\documentclass[twocolumn,10pt]{article}
\usepackage{epsfig}
\makeatletter
\renewcommand{\section}{\@startsection
   {section}
   {1}
   {\z@}
   {-10mm}
   {2mm}
   {\normalfont\normalsize\bfseries}}
\renewcommand{\subsection}{\@startsection
   {subsection}
   {2}
   {\z@}
   {-5mm}
   {1mm}
   {\normalfont\normalsize\itshape}}
\makeatother

\def\rhs{r_{\rm hs}}

\newcommand{\beq}{\begin{eqnarray}}
\newcommand{\eeq}{\end{eqnarray}}
\newbox\grsign \setbox\grsign=\hbox{$>$} \newdimen\grdimen \grdimen=\ht\grsign
\newbox\simlessbox \newbox\simgreatbox \newbox\simpropbox
\setbox\simgreatbox=\hbox{\raise.5ex\hbox{$>$}\llap
     {\lower.5ex\hbox{$\sim$}}}\ht1=\grdimen\dp1=0pt
\setbox\simlessbox=\hbox{\raise.5ex\hbox{$<$}\llap
     {\lower.5ex\hbox{$\sim$}}}\ht2=\grdimen\dp2=0pt
\setbox\simpropbox=\hbox{\raise.5ex\hbox{$\propto$}\llap
     {\lower.5ex\hbox{$\sim$}}}\ht2=\grdimen\dp2=0pt
\def\simgreat{\mathrel{\copy\simgreatbox}}

\begin{document}
\twocolumn[
{\Large \bf \textsf{Unification of Spectral States of Accreting Black Holes}
\par}
\vspace*{2mm}
Juri Poutanen$^{1,2}$ and Paolo Coppi$^3$ \par \vspace*{2mm}
{\small $^1$ Stockholm Observatory, SE-13336 Saltsj\"obaden, Sweden \\
$^2$ Astronomical Observatory, Box 515, SE-75120 Uppsala, Sweden \\
$^3$ Astronomy Department, Yale University, P.O.Box 208101, New Haven, 
CT 062520-8101, USA\\[1mm]}
\vspace{15mm}
]

\section*{Abstract}

{\small
Several galactic black holes show transitions between spectral states.
The nature of these transitions is not fully understood yet.
None of the dynamical accretion disk models can fully describe spectral
transitions. In this paper we present a unifying
radiation transfer model that can fit the
spectral data in both states. Since Cyg X-1 has the best available data, we
focus here on modeling this object.
We fit individual broad-band (from 1 keV up to 4 MeV)
spectral data for the ``hard'' and ``soft'' states of Cyg X-1  using an emission
model where a central Comptonizing corona/cloud is illuminated by the soft
photon emission from a cold disk. 
We assume that the energy is injected to the corona by two channels: 
a non-thermal one that injects energetic ($>$ MeV)
electrons into the coronal region, and a thermal one that heats injected
and ambient electrons once they cool sufficiently to form a
Maxwellian distribution, i.e., we consider a hybrid thermal/non-thermal
model.  The process of photon-photon pair production
is included in the model, and the number of pairs produced in the
coronal region can be substantial.

Using simple scaling laws for the luminosity of the cold disk, the
thermal dissipation/heating rate in the corona,
and the rate of energy injection from a
non-thermal source, all as functions of radius of the corona,
we explain the hard-to-soft transition as the result of a decrease in the
size of the corona and the inner radius of the cold disk by
a factor $\sim$ 5. For the case of Cyg X-1, we show that the bolometric
luminosity of the source (mass accretion rate) does not change
significantly during the transition, and thus the transition is
probably the result of a disk instability.
}

\section{Observations of Galactic Black Holes}

Galactic black holes radiate X-rays and $\gamma$-rays in one
of several spectral states.
Recently, it has become clear that in the soft state,
the power-law spectrum (with typical energy spectral index $\alpha
\sim 1.5$) extends to photon energies $\sim m_ec^2$ without
any obvious break [1-3]. The black hole
spectra in the hard state instead show a cutoff at $\sim$ 100 keV [3-5] 
and can be fit quite well with
thermal Comptonization models up to energies $\sim 300$ keV. However,
there is evidence in the   COMPTEL and BATSE data
that the hard state spectrum of Cyg X-1 show high energy excess at $\simgreat$ 
500 keV
[6,7].
This excess can be interpreted as a signature of non-thermal
electrons in the X/$\gamma$-ray source [8,9]. 
This excess can also be explained
in terms of a multi-zone models where thermal electrons have
significantly different temperatures [7,10,11]. 

While the data in any one state can be fit quite
successfully by one of the models mentioned above, none of the
models proposed thus far can fit data from both states.
Esin et al. [12] explain
the spectral transitions in terms of the advection dominated disks.
Their prediction that
electron temperature in the soft state should be smaller than that
in the hard state (because of the significant increase in the number of
ambient soft photons) is in perfect agreement with our findings. 
However, their model does not explain X/$\gamma$-ray spectra that extend
unbroken to $\simgreat$ 1 MeV, since they do not consider non-thermal 
particles. 
The soft state power-laws were also explained 
in terms of bulk Comptonization in a converging flow [13].
This model would require, however, a significant increase in the
accretion rate to account for the significant decrease in the radiative
efficiency of the spherical
converging flow, which does not appear to be the case, at least
for the most recent hard-soft transition of Cyg X-1. 
The observed signatures of Compton reflection with an iron edge
smeared out by relativistic effects [14] and the possible extend of the 
power-law spectrum up to $\sim$ 10 MeV [15] 
would be difficult to explain in such a scenario.

Since fully dynamical (accretion disk) models are still highly uncertain,
in our model, we instead concentrate  on the radiative
processes that lead to the formation of the observed broad-band
spectra of accreting black holes.
We propose a ``radiative'' model (as opposite to dynamical)
which is based on the hybrid thermal/non-thermal pair model
and can explain both the  soft and hard spectra
of galactic black holes in a  unified way.

\section{Hybrid Thermal/Non-Thermal Pair Model (HPM)}

The model is based on the pair plasma code of Coppi [16]. 
The model incorporates the following processes:
Compton scattering (using exact the Klein-Nishina cross-section
and redistribution function),
pair-production and annihilation, Coulomb scattering, and bremsstrahlung.
Compton reflection from the cold matter of an arbitrary ionization state
is accounted for using angular dependent Green functions [17]. 
Our model is a single-zone model. (All particle and distributions
are assumed to be isotropic and homogeneous in the coronal region.)
The input parameters of the model are:
(i) the thermal compactness, $l_{th}=L_{th}/r_c \cdot\sigma_T/(m_ec^3)$
which characterizes the heating rate of electrons (pairs);
(ii) the analogous non-thermal compactness, $l_{nth}$, which
characterizes the  rate of injection of relativistic electrons,
(iii) the soft photon compactness, $l_s$, which represents the fraction
of the cold disk luminosity that enters the X/gamma-ray source (corona),
(iv) $\Gamma_{inj},$ the power-law index of the non-thermal electron
injection spectrum,
(v) $\tau_p$ the proton (Thomson) optical depth
(ie., the optical depth due to background electrons), and (vi) $T_{bb}$, 
temperature of the soft black body radiation.
Here $r_c$ is the radius of the corona.

Instead of  introducing (ad hoc) a number
of zones with different temperatures, we propose the model where electrons
are not thermal, but the electron distribution is computed self-consistently
balancing electron cooling (by Compton scattering and Coulomb interactions),
heating (thermal energy source), and acceleration (non-thermal energy source).
Both the electron and positron energy distributions 
are assumed to consist of a Maxwellian distribution
of arbitrary temperature plus a non-thermal tail of arbitrary
shape (which is again solved for self-consistently; in general, 
it is {\it not} a power law).
The code used in these simulations is incorporated into XSPEC. 
Although it is simple, this one-zone
model appears to be able fit black hole spectra very well.

\section{Spectral States}

\begin{figure}[h]
\epsfig{figure=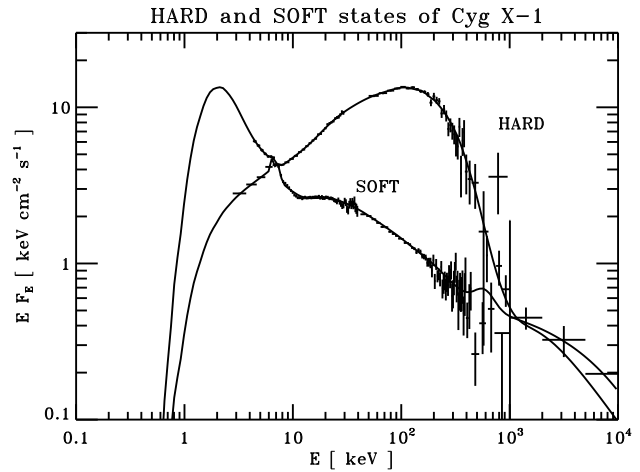,width=9cm}
\caption{\label{fit}\small 
Best fits to the two spectral states of Cyg X-1 using
the hybrid thermal/non-thermal pair model. The hard state data are simultaneous 
observations by Ginga, OSSE, and COMPTEL (June 1991). 
The best fit model has 
the following parameters $l_{nth}=1.3$, $l_{s}=1$ (frozen), $l_{th}=11$,
$\tau_p=1.47$, $T_{bb}=0.13$ keV (f), and Compton reflection amplitude $R=0.3$, give  a
$\chi^2$/dof=56/78.
The soft state data are from the XTE/PCA and OSSE
(June 1996). The model spectrum corresponds to the 
fit parameters: $l_{nth}=5$ (f), $l_{s}=34$, $l_{th}=3$,
$\tau_p=0.23$ (f), $T_{bb}=0.39$ keV, $R=1.1$, giving a 
$\chi^2$/dof=180/169.
In both cases the power-law index for non-thermal
electron injection was fixed at
$\Gamma_{inj}=2.4$. Spectra cut off at low energies due to interstellar 
absorption with column density $N_H=0.6\cdot 10^{22}$ cm$^{-2}$.  
}
\end{figure}

We fit the hybrid pair model to the broad-band data of Cyg X-1 in both
hard and soft states  using XSPEC.
The simultaneous hard-state observations by Ginga, OSSE and COMPTEL [4,6] 
are analyzed together. 
For the soft state, publically available  XTE/PCA and OSSE data from
June 1996 are analyzed.
The data together with the best fit models are
presented in Figure~\ref{fit}.

We interpret the spectral transition as a result of decrease of the
inner radius of the cold accretion disk (and corresponding
decrease in the size of the coronal region).
In the hard state,
the inner radius of the cold disk is far away from the black hole
(at $\sim 30-50 GM/c^2$). Almost all the energy is dissipated thermally in
the hot central cloud/corona.
In addition, we assume that there is a source of non-thermal electrons
(an ``accelerator'') situated close to the black hole which pumps
energetic electrons into the corona.  Its contribution  to the total
luminosity is of order of 10\%. The X/gamma-ray source is photon
starved since the luminosity of the outer cooler part of the accretion flow
is small and the covering factor of the hot corona is quite small too
(see, e.g., [18]).  
Amplitude of the Compton reflection is relatively small, $R\sim 0.3$ [4,19]. 
Under these conditions, electrons have an almost Maxwellian distribution with
a weak power-law tail.
This results in the hard spectrum produced by thermal Comptonization with
a small contribution from the non-thermal tail at energies above $\sim m_ec^2$.

In the soft state, the situation is very different.
The optically thick cool disc moves inward and receives the
majority of the dissipated energy. The thermal energy dissipation in the
corona is now a factor of $\sim$ 10 lower than the soft luminosity (from
the accretion disk) entering the corona. The non-thermal electron luminosity
(which we take to be constant) is larger than the thermal
dissipation rate. The electron distribution can be represented by a thermal
distribution with a lower temperature (relative to the hard state)
of $\sim$ 20 keV and significant power-law tail. Most of the power is
now in non-thermal electrons.
The resulting spectrum is a black body from
the cold accretion disk, followed by a ``soft excess'' due to
Comptonization by the thermal population of electrons (pairs),  and
then a power-law (Comptonization by non-thermal electrons)
which extends up to $\sim kT_{bb}\gamma_{max}^2$
(here $\gamma_{max}$ is the maximum electron energy). Note that since
most of the observed X-rays are produced in {\it one} scattering off
non-thermal electrons, we expect no significant time lags between
hard and soft photon energies (as would be the case for thermal
Comptonization).

\section{Spectral Transitions}

\begin{figure}[h]
\epsfig{figure=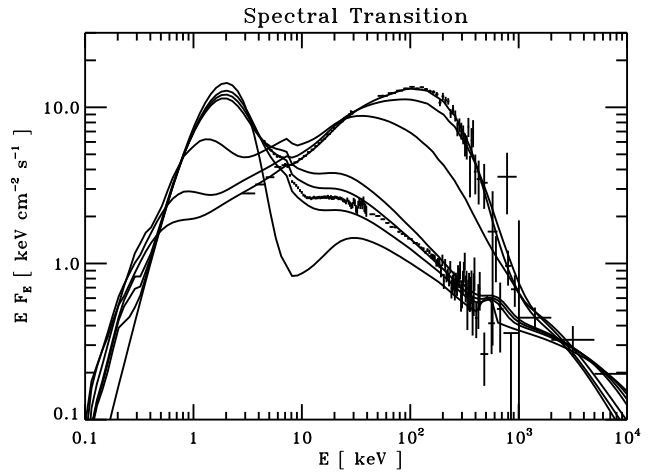,width=9cm}
\caption{\label{states}\small 
Simulations of the spectral transitions by the HPM.
The starting point is the best fit to the hard state data ($r=6$).
Using the simple scaling laws (see \S~4) for $l_s, l_{th}, l_{nth}$,
we obtain the sequence of spectra covering  the transition between
the hard and soft states as a function of the inner disk radius, $r$.
From top to bottom (at 100 keV), the curves correspond to the
inner disk radius of
6, 4, 2, 1.2, 1.15, 1.1, 1.0, respectively.
The spectra at $r\sim 1.1-1.15$ are quite close to the observed
soft state spectrum.
}
\end{figure}

We model the changes in the source during the spectral hard-to-soft
by using simple
scaling laws for the soft compactness, $l_s$, thermal compactness, $l_{th}$,
and non-thermal compactness, $l_{nth}$, as well as for proton optical
depth, $\tau_p$, and
covering fraction of the cold disk around hot corona, $R$.
Assume that thermal dissipation in the corona vanishes when 
inner radius of the cold accretion disk $r=1$ (arbitrary units), 
and the radius of the cold disk in the hard state is $\rhs>1$. 
Now, further assume that the radius of the corona is constant in units
of the inner radius of the accretion disk (which is, probably, 
not exactly the case, see [18]). 
The sum of soft luminosity from the disk, $L_s\propto 1/r$, and 
the thermal dissipation rate in the corona, $L_{th}\propto 1-1/r$, 
should remain approximately constant during transition 
(required to fit the data).
A decrease in radius $r$ during the transition thus gives
the following dependences on $r$ for the model parameters:
\beq
l_s(r)     & = & l_s^{\rm hs} \left(\frac{\rhs}{r}\right)^2 ,\nonumber \\
l_{th}(r)  & = & l_{th}^{\rm hs}\frac{\rhs}{r} \frac{1-1/r}{1-1/\rhs}, 
\nonumber \\
l_{nth}(r) & = & l_{nth}^{\rm hs} \frac{\rhs}{r} . \nonumber 
\eeq
The quadratic dependence of $l_s$ on $r$ follows from the fact that
$L_s\propto 1/r$ and compactness $l\propto L/r$.
The thermal luminosity decreases when inner radius of the disk decreases, while
the thermal compactness parameter $l_{th}$ grows at large $r$
and the rapidly decreases when $r$ is close to 1.
We assume here that proton optical depth follows the (ad hoc) relation
$\tau_p=\tau_p^{\rm hs} r/\rhs.$ (The exact form does not appear to be
very important.) Since the amplitude of the Compton
reflection is higher in the soft state, we take $R=R_{\rm hs}(r/\rhs)^{0.8}$.
The temperature of the soft photon black body can be
computed, of course, from the soft luminosity and the size of accretion disk
$T_{bb}=T_{bb}^{\rm hs}(r/\rhs)^{-3/4}$.
The whole sequence of  spectra for different $r$ between $r=\rhs=6$ and
$r=1$ is shown in Figure~\ref{states}. The spectral shape changes dramatically
in a very narrow interval of radii between $r=2$ and $r=1$, where the thermal
energy dissipation in the corona decreases significantly.

\section{Summary}

We present an emission model that unifies the hard and  the soft state of
accreting black holes. We show that even when the total source
luminosity (the mass accretion rate) is constant, the
broad-band source spectrum can significantly
change its shape simply due to redistribution of the energy release between
the cold outer disk and a hot central corona. (Note that the
energy injection rate into non-thermal particles required to explain
the soft state spectrum as well as the MeV tail in hard state
is also assumed to remain constant during the transition.)

When inner radius of the cold
accretion disk is large (tens of $GM/c^2$), most of the power
is dissipated in a central cloud/corona-like structure
(the inner hot accretion disk).
The Comptonization by mostly thermal electrons produces a spectrum which
cuts off at $\sim$ 100 keV, and a weak power-law tail (due to non-thermal
population of electrons) extending up to MeV energies. This is the hard state.

In the soft state, most of the power is dissipated in the cold disk, and
emerges in form of blackbody-like spectrum. The thermal energy supply to the
corona becomes negligible. The emerging spectrum thus  consists of a
black body, a power-law (produced by non-thermal Comptonization) extending
up to MeV energies, and a ``soft excess'' at a few keV (produced by
Comptonization in a cool thermal plasma with temperature of order
$\sim 20$ keV). Note that the temperature and density
of the thermal plasma are not free parameters and are
determined self-consistently in the model.
In both  states, Compton reflection bump changes the intrinsic spectrum
significantly from $\sim$ 5 to 100 keV.

This research was partially supported by grants from the
Swedish Natural Science Research Council and   
the Anna-Greta and Holger Grafoord's Fund. 
This study has made use of data obtained from the HEASARC, provided 
by NASA's Goddard Space Flight Center. 
We would like to thank 
Boris Stern, Roland Svensson, Andrzej Zdziarski and Marek Gierli\'nski 
for useful discussions. 

\section{References}
\small

\begin{enumerate}
\item
   Grove, J.~E., Kroeger, R.~A. and Strickman, M.~S., 
     "The Transparent Universe", Proc. 2nd INTEGRAL workshop,
     (ESA SP-382, 1997), p.197.  
\item 
   Grove, J.~E., {\it et al.}, 
   Proc. 4th Compton Symposium (AIP, New York 1997), in press. 
\item
   Phlips, B.~F., {\it et al.}, ApJ, {\bf 465}, 907 (1996).  
\item
   Gierli\'nski, M., {\it et al.},  MNRAS, {\bf 288}, 958 (1997). 
\item 
  Zdziarski, A.~A., Johnson, W.~N., Poutanen, J., Magdziarz, P. and
  Gierli\'nski, M.,   ``The Transparent Universe'', Proc. 2nd INTEGRAL workshop,
  (ESA SP-382 1997), p.373 (astro-ph/961210).
\item 
McConnell, M.~L.,  {\it et al.}, ApJ, {\bf 424}, 933 (1994). 
\item
Ling, J.~C., {\it et al.}, ApJ, {\bf 484}, 375 (1997). 
\item 
Crider, A., Liang, E.~P., Smith, I.~A., Lin, D. and Kusunose, M., 
Proc. 4th Compton Symposium (AIP, New York 1997), in press (astro-ph/9707006). 
\item
Liang, E.~P. and Narayan, R.,
Proc. 4th Compton Symposium (AIP, New York 1997), in press. 
\item 
Liang, E.~P., ApJ, {\bf 367}, 470 (1991). 
\item 
Moskalenko, I.~V., Collmar, W. and Sch\"onfelder, V., 
Proc. 4th Compton Symposium (AIP, New York 1997), in press (astro-ph/9709179). 
\item 
Esin, A.~A., McClintock, J.~E. and Narayan, R., ApJ, {\bf 489}, 865 (1997)
\item
Titarchuk, L., Mastichiadis, A. and Kylafis, N.~D., ApJ, {\bf 487}, 834 (1997). 
\item 
   Gierli\'nski, M., {\it et al.},
Proc. 4th Compton Symposium (AIP, New York 1997), in press (astro-ph/9707213). 
\item 
 Iyudin, A.~F., private communication. 
\item 
Coppi, P.~S., MNRAS, {\bf 258}, 657 (1992).
\item 
Magdziarz, P. and Zdziarski, A.~A., MNRAS, {\bf 273}, 837 (1995). 
\item
Poutanen, J., Krolik, J.~H. and Ryde, F., MNRAS, {\bf 292}, L21 (1997) 
\item 
Ebisawa, K., {\it et al.}, ApJ, {\bf 467}, 419 (1996). 

\end{enumerate}

\end{document}